\documentclass[aps,prd,reprint,amsmath,amssymb,longbibliography,nofootinbib]{revtex4-2}
\usepackage{graphicx}
\usepackage{bm}
\usepackage{amsmath}
\usepackage{amsfonts}
\usepackage{amssymb}
\usepackage{color}
\usepackage{lipsum}
\usepackage{dcolumn}
\usepackage{listings}
\usepackage[dvipsnames]{xcolor}
\usepackage{hyperref}
\usepackage{multirow}

\usepackage{enumitem}

\usepackage[capitalise]{cleveref}
\usepackage[activate={true,nocompatibility},final,kerning=true,factor=1100,stretch=10,shrink=10]{microtype}
\usepackage{academicons}
\usepackage{fontawesome5}
\definecolor{orcidlogocol}{rgb}{0.65, 0.807, 0.223}
\newcommand{\orcid}[1]{$\,$\href{https://orcid.org/#1}{\textcolor{orcidlogocol}{\faOrcid}}}

\newcommand{\Planck}{\textit{Planck}}

\newcommand{\lcdm}{$\Lambda$CDM}

\newcommand{\beq}{\begin{equation}}
\newcommand{\eeq}{\end{equation}}

\def\ee{\end{equation}}
\def\bea{\begin{eqnarray}}
\def\eea{\end{eqnarray}}

\newcommand{\pact}{\textsf{P-ACT}}
\newcommand{\pactlb}{\textsf{P-ACT-LB}}

\lstdefinestyle{yamlstyle}{
  backgroundcolor=\color{gray!10},
  basicstyle=\ttfamily\small,
  numbers=left,
  numberstyle=\tiny\color{gray},
  breaklines=true,
  showstringspaces=false,
  morekeywords=[2]{true,false}, 
  keywordstyle=[2]\color{blue},
  literate=
    {:}{{{\color{red}{:}}}}1
    {0}{{{\color{blue}0}}}1
    {1}{{{\color{blue}1}}}1
    {2}{{{\color{blue}2}}}1
    {3}{{{\color{blue}3}}}1
    {4}{{{\color{blue}4}}}1
    {5}{{{\color{blue}5}}}1
    {6}{{{\color{blue}6}}}1
    {7}{{{\color{blue}7}}}1
    {8}{{{\color{blue}8}}}1
    {9}{{{\color{blue}9}}}1
}

\begin{document}
\title{Ultra-light axion constraints from Planck and ACT: the role of nonlinear modelling}
\author{Lauren Gaughan \orcid{0009-0004-8338-0180}}
\email{lauren.gaughan@nottingham.ac.uk}
\affiliation{University of Nottingham}
\author{Anne M. Green\orcid{0000-0002-7135-1671}}
\email{anne.green@nottingham.ac.uk}
\affiliation{University of Nottingham}
\author{Adam Moss\orcid{0000-0002-7245-7670}}
\email{adam.moss@nottingham.ac.uk}
\affiliation{University of Nottingham}
\date{\today}

\begin{abstract}
We study how constraints on the abundance of ultralight axions (ULAs) from cosmic microwave background (CMB) data depend on their nonlinear modelling. We focus on the axion mass range $10^{-25} \leq m/\rm{eV} \leq 10^{-23}$, where the axion Jeans scale falls in the quasi-linear regime probed by CMB lensing, making constraints highly sensitive to the choice of nonlinear prescription. We show that the inferred constraints depend significantly on the choice of nonlinear model, which must therefore be treated carefully. Performing Markov Chain Monte Carlo (MCMC) analyses with \Planck\, 2018, ACT DR6 and DESI DR2 BAO data, we find naive nonlinear modelling of non-cold matter can produce an artificial preference for a subdominant ULA dark matter component with mass $m \approx 10^{-24}\,$eV. This arises from a lensing-like enhancement of the CMB power spectrum. 
\end{abstract}

\maketitle

\section{Introduction}

The $\Lambda$CDM model has had great success in describing the evolution of the Universe \cite{Planck:2018vyg}. Several questions, however, still persist, including the nature of dark matter and dark energy. While a wide range of candidates have been proposed, ultralight axions (ULAs) \cite{Marsh:2015xka} have gained popularity in recent years. The QCD axion was originally proposed to solve the strong-CP problem in quantum chromodynamics. Lighter axions arise in extensions to the standard model, such as the string theory axiverse \cite{Arvanitaki:2009fg,Svrcek:2006yi} in which pseudo-Nambu-Goldstone bosons (or axion-like particles) are produced via symmetry breaking. 

The behaviour of ULAs depends on the relationship between the axion mass, $m$, and the Hubble parameter, $H$~\cite{Hlozek:2014lca}. When $H > m$, the axion field is frozen and behaves like dark energy. As the Universe expands, $H$ decreases and once $m \sim H$, the field begins to oscillate and subsequently behaves like cold dark matter (CDM). Crucially, the mass corresponding to the Hubble rate at matter-radiation equality, $H_{\rm{eq}} \sim 10^{-28}\,$eV, provides the distinction between lighter ULAs ($m \lesssim 10^{-28}\,$eV), which behave like dark energy at late times, and heavier ULAs, which behave like dark matter. The full ULA  mass range is $10^{-33} \lesssim m/\rm{eV} \lesssim 10^{-18}$, bounded below by the current Hubble rate $H_0$ (lighter axions are still frozen today) and above by the scale at which wave-like effects become astrophysically negligible.

ULAs are a specific realisation of fuzzy dark matter, a class of dark matter candidates whose wave-like behaviour manifests on astrophysical scales. Due to their low mass, ULAs have large de Broglie wavelengths, $\lambda = h/(mv)$, which can extend to galactic or larger scales. On scales larger than the de Broglie wavelength, ULAs are almost indistinguishable from CDM. On smaller scales, quantum pressure opposes gravity, preventing collapse, and instead the density perturbation oscillates. If axions make up a non-negligible fraction of the dark matter, the matter power spectrum is suppressed~\cite{Hu:2000ke}. Standard CDM, however, does cluster on these small scales, so mixed dark matter (MDM) cosmologies---with both fuzzy and cold components---require more careful treatment. See Refs.~\cite{Marsh:2015xka,Hui:2021tkt} for more detailed reviews.

Constraints on the abundance of ULAs are usually quoted in terms of the axion fraction, defined as the fraction of the total dark matter that is axions:
\begin{equation}
    f_{\rm{ax}} \equiv \frac{\Omega_{\rm{ax}}}{\Omega_{\rm{D}}} \,,
\end{equation}
where $\Omega_{\rm{ax}}$ and $\Omega_{\rm{D}}$ are the axion and total dark matter density parameters respectively.
Cosmic microwave background (CMB) data constrain ULAs because the suppression of small-scale structure growth by ULAs alters the angular power spectrum. Hlozek et al.~\cite{Hlozek:2014lca, Hlozek:2017zzf} used the standard effective fluid approximation (EFA) in \texttt{axionCAMB} and \Planck{} 2015 data to place the constraint $f_{\rm{ax}} \leq 0.05$ at 95\% confidence~\footnote{All subsequent limits are also quoted at 95\% confidence.} in the mass range $10^{-32} \leq m/\rm{eV} \leq 10^{-25.5}$.  Rogers et al.~\cite{Rogers:2023ezo} found stronger constraints at lower masses using \Planck{} 2018, Atacama Cosmology Telescope (ACT) DR4 and South Pole Telescope (SPT-3G) CMB data: $f_{\rm{ax}} < 0.1$ for $m \lesssim 10^{-26}\,\rm{eV}$, tightening to $f_{\rm{ax}} < 0.01$ for $10^{-30} \leq m/\rm{eV} \leq 10^{-28}$. ULAs with $m \gtrsim 10^{-24}\,\rm{eV}$ are effectively degenerate with standard CDM on CMB length scales; thus the CMB alone does not constrain $\Omega_{\rm{ax}}h^{2}$, and large axion fractions remain allowed at these masses.

The rapid oscillations of the ULA field become computationally prohibitive because small timesteps are required to accurately track the evolution. The standard EFA~\cite{Hu:2000ke, Hlozek:2014lca} addresses this by averaging over the oscillations and treating the axion field as a fluid. In Ref.~\cite{Moss:2025ymr} we presented an implementation of the improved Passaglia and Hu (PH) EFA~\cite{Passaglia:2022bcr} in \texttt{AxiCAMB} (publicly available at \url{https://github.com/adammoss/AxiCAMB}). The PH EFA factors out the rapid axion-field oscillations by introducing auxiliary fields, and calibrates the effective fluid equation of state and sound speed, providing a more reliable approximation of the axion behaviour. Using \Planck{} PR4 and DESI DR1 baryon acoustic oscillation (BAO) data, we obtained improved constraints compared to the standard EFA implementation in Ref.~\cite{Rogers:2023ezo}: $f_{\rm{ax}} < 0.044$ for $10^{-26}\,\rm{eV}$, tightening to $f_{\rm{ax}} < 0.0082$ for $10^{-28}\,\rm{eV}$.

The recent ACT DR6 analysis~\cite{AtacamaCosmologyTelescope:2025nti} used the standard EFA in \texttt{axionCAMB} and also the \texttt{axionEMU} emulator, a machine-learning emulator trained on \texttt{axionCAMB} outputs, in the mass range $10^{-28} \leq m/\rm{eV} \leq 10^{-25}$. Combining \Planck{} 2018 and ACT CMB data, they found tighter constraints for $m \geq 10^{-26}\,\rm{eV}$ than from \Planck{} alone. However, they noted that for $m \geq 10^{-25}\,\rm{eV}$ more detailed nonlinear modelling is required, and deferred this to future work.

While the CMB alone is limited by the angular scales it probes, small-scale structure data can significantly tighten the constraints. Lyman-$\alpha$ forest data rule out $f_{\rm{ax}} = 1$ for $m < 2\times 10^{-20}\,\rm{eV}$~\cite{Rogers:2020ltq}, though ULAs in this mass range can still make up a subdominant component at the ${\cal O}(10)\%$ level. Winch et al.~\cite{Winch:2024mrt} find $f_{\rm{ax}} < 0.28$ from HST UV luminosity function (UVLF) data alone and $f_{\rm{ax}} < 0.15$ from HST UVLF combined with \Planck{} 2018 for $m = 10^{-24}\,$eV.

The reliability of small-scale structure constraints depends crucially on the nonlinear modelling. CMB experiments that measure gravitational lensing are also sensitive to nonlinear scales, and this sensitivity increases with the precision of the lensing reconstruction. In this work, we investigate the effect of nonlinear modelling for axions in the mass range $10^{-25} \leq m/\rm{eV} \leq 10^{-23}$. For masses in this range, the axion Jeans cutoff lies in the quasi-linear regime, $k \sim (0.1$--$1) \, h$ Mpc$^{-1}$, that contribute significantly to
CMB lensing. The CMB-lensing kernel is broad, with sensitivity over
$z\sim0.5$--$5$ and a peak near $z\simeq2$. For lighter masses, the linear suppression of power is already large on CMB scales, so nonlinear corrections are a secondary effect. For heavier masses, the axion becomes nearly degenerate with CDM in CMB observables. 

This mass window is also well motivated phenomenologically. Rogers et al.~\cite{Rogers:2023ezo} suggested that axions with a mass of $10^{-25}\,$eV can reduce the $S_{8}$ tension between CMB and galaxy clustering data. In addition, the formation of axion soliton cores may alleviate the cusp-core problem \cite{Hui:2016ltb}. Finally, upcoming CMB lensing measurements from experiments such as Simons Observatory will probe the heavier end of this mass window, particularly on nonlinear scales.

We integrate the \texttt{axionHMcode} \cite{Vogt:2022bwy} nonlinear model for ULAs into the Boltzmann solver \texttt{AxiCAMB}, including both the basic and DOME-calibrated \cite{Dome:2024hzq} prescriptions.  These nonlinear models modify the matter power spectrum and halo mass function, and hence affect the inferred constraints on the axion fraction, $f_{\rm{ax}}$, and physical density, $\Omega_{\rm{ax}}h^{2}$. As shown in Tab.~\ref{tab:chi2_f_ax}, the effective $\Delta\chi^{2}$ relative to $\Lambda$CDM can indicate a preference of up to $|\Delta \chi^2| \sim 4.56$ for the presence of axions, depending on the nonlinear model.

In Sec.~\ref{sec:efa}, we provide a brief summary of the standard and PH EFAs, emphasizing the improvements provided by the PH EFA. 
Section \ref{sec:nonlinear} details current nonlinear modelling within Boltzmann codes and the differences between $\Lambda$CDM and ULA cosmologies, and then discusses the MDM halo model implementation within \texttt{axionHMcode} used to calculate ULA nonlinear corrections. 
In Sec.~\ref{sec:method} we outline the CMB, lensing and baryon acoustic oscillation (BAO) data and settings we use within \texttt{Cobaya} to perform Markov Chain Monte Carlo (MCMC) analysis. Section \ref{sec:results} discusses the results of the MCMC calculations. 
Finally, in Sec.~\ref{sec:discussion} we conclude with a discussion of the results and future directions.

\section{Effective Fluid Approximation} \label{sec:efa}

 We treat the axion as an effective fluid, using the implementation of Passaglia and Hu (PH) \cite{Passaglia:2022bcr} presented in our previous work \cite{Moss:2025ymr}; here we provide a brief overview of the model.

The evolution of the axion field, $\phi (\vec{x}, t)$, is described by the Klein-Gordon (KG) equation:
\begin{equation}
    \Box \phi(\vec{x}, t) = \frac{{\rm d} V(\phi(\vec{x}, t))}{{\rm d} \phi} 
    \approx m^2 \phi(\vec{x}, t)\,,
\end{equation}
where $m$ is the mass of the axion. As is standard, we assume the axion field is sufficiently close to the minimum that the cosine potential can be approximated by its quadratic form $V(\phi) \approx \frac{1}{2}m^{2}\phi^{2}$.

In a Friedmann-Lema\^itre-Robertson-Walker (FLRW) cosmology, the background equation of motion for the axion field is given by
\begin{equation}
    \ddot{\phi} + 2\mathcal{H}\dot{\phi} + a^{2}m^{2}\phi = 0 \,,
    \label{eq: background axion eom}
\end{equation}
where overdots are derivatives with respect to conformal time, and $\mathcal{H} = aH$ is the conformal Hubble parameter. At early times, while $m \ll H$, the field is frozen and behaves as dark energy, with equation of state parameter $w = -1$. As the Universe evolves, $H$ decreases, and once $m \sim H$ the field begins to oscillate. 
At late times, when $m \gg H$, the equation of state averages to zero, $\langle w \rangle = 0$, and the field behaves as CDM. These rapid oscillations become much shorter than cosmological timescales, requiring prohibitively small time steps for direct numerical integration in Boltzmann codes.

To avoid these numerical issues, an effective fluid approximation is used. The standard EFA evolves the full axion field KG equation until oscillations begin. At a chosen scale factor $a_{*}$, set by $m \approx 3H(a_{*})$ in the original \texttt{axionCAMB} implementation, the field is switched to a fluid description. This choice captures late-time evolution accurately while allowing sufficient time for numerical transients to decay. The fluid variables ($\rho_{\rm{ax}}, \delta_{\rm{ax}}, u_{\rm{ax}}$) are initialized at $a_{*}$ using values from the KG equation. After this transition, the axion perturbations are treated as a pressureless fluid with equation of state $w_{\rm{ax}}=0$ and effective sound speed $c_{\rm s}$:
\begin{equation}
    c^{2}_{\rm{s}} = \frac{k^{2}}{4m^{2}a^{2} + k^{2}} \,.
\end{equation}
For $k \ll 2ma$, this reduces to $c_{\rm{s}}^{2} \approx k^{2}/(4m^{2}a^{2})$. This non-zero sound speed leads to a Jeans scale, below which axion density perturbation growth is suppressed. At late times, the EFA captures the average effect of the rapid oscillations, whilst not directly solving the full KG equation. For axion masses $m \lesssim 10^{-24}\,\rm{eV}$, this approximation breaks down because the transition from the frozen state to oscillations occurs near or after recombination, so the fluid approximation is applied too early.

The PH EFA \cite{Passaglia:2022bcr} addresses these limitations by implementing matching conditions between the field and fluid at a switch time set by $m/H$, and also using a corrected expression for the sound speed.
It decomposes the axion field into two auxiliary fields, $\varphi_{\rm{c}}$ and $\varphi_{\rm{s}}$:
\begin{equation}
    \phi(\tau) = \varphi_{\rm{c}}(\tau) \cos [\tau - \tau_{*}] + \varphi_{\rm{s}}(\tau) \sin [\tau - \tau_{*}],
\end{equation}
where $\tau=mt$ is a dimensionless time variable, and $\tau_{*}$ is the switch time, which occurs after oscillations have begun. The perturbations of the axion field $\delta\phi$ are treated similarly and decomposed into $\delta\varphi_{\rm{c}}$ and $\delta\varphi_{\rm{s}}$. This decomposition factors out the rapid oscillations. In our previous work \cite{Moss:2025ymr}, we found that $m/H_{*} = 50$ provides high accuracy at late times whilst still capturing the early oscillatory phase effectively. 

The matching conditions allow the ratio of the second and first derivatives of the auxiliary fields $\varphi_{\rm{c,s}}$ to be determined at the switch time: $\left.\frac{\varphi''_{\rm{c,s}}}{\varphi'_{\rm{c,s}}}\right|_{*} = D$, where
\begin{equation}
\label{eq:constraints}
D \equiv \left.-\frac{1}{2} \frac{\langle H \rangle}{m} \left(3 - \frac{m}{\langle H\rangle ^3}\frac{d\langle H\rangle^2}{d \tau}\right)\right|_{*} \,,
\end{equation}
with $\langle H \rangle$ the Hubble parameter averaged over axion oscillation cycles. In practice, the effective Hubble rate at the time of the switch is used. Similarly, to suppress oscillations, the matching condition for the axion perturbations takes the same form, with $\left.\frac{\delta\varphi''_{\rm{c,s}}}{\delta\varphi'_{\rm{c,s}}}\right|_{*} = D$. Solving these conditions initialises the auxiliary variables for the background and perturbations at the switch time, providing the initial conditions for the effective fluid variables ($\rho^{\rm{ef}}_{\rm{ax}}, \delta^{\rm{ef}}_{\rm{ax}}, u^{\rm{ef}}_{\rm{ax}}$).

Before matching, the axion perturbations are evolved as a scalar field, and afterwards as a fluid with the calibrated sound speed\footnote{A typographical error appeared in the printed expression for this sound speed in Ref.~\cite{Moss:2025ymr}: the first term was written without squaring the field sound speed. The numerical implementation used for the results in that work used the expression shown here, so the parameter constraints and likelihood results are unchanged.}
\begin{equation}
    c_{\rm{s}}^2 = \left(\frac{k}{a m}\right)^{-2} \left( \sqrt{1+\left(\frac{k}{a m} \right)^2} -1 \right)^{2} + \frac{5}{4} \left(\frac{m}{H}\right)^{-2}\,,
\end{equation}
where the second term is a late-time correction to the sound speed at finite $(m/H)^{-1}$.
 Ref.~\cite{Passaglia:2022bcr} computed $\delta P^{\rm ef}/\delta \rho^{\rm ef}$, finding that the effective sound speed approaches a nonzero value as $k/am \rightarrow 0$, with a correction of order $(m/H)^{-2}$ and coefficient $\sim 5/4$. This calibration avoids the spurious features in $P(k)$ that the standard EFA introduces, particularly at $k \sim$ few $\times\, 0.01$~$h$ Mpc$^{-1}$ for $m = (10^{-25}$--$10^{-24}) \,$eV. See Ref.~\cite{Moss:2025ymr} for full details and validation.
 
\section{Nonlinear Modelling} \label{sec:nonlinear}

Nonlinear corrections to the matter power spectrum, $P(k)$, affect CMB observables via gravitational lensing from large-scale structure, which smooths the shape of the angular power spectra. The lensed CMB power spectra $C_{\ell}^{TT}$ and $C_{\ell}^{EE}$, as well as
the lensing potential power spectrum $C_L^{\phi\phi}$, depend on a line-of-sight
projection of the matter power spectrum. For CMB lensing, the
redshift sensitivity is broad, extending over $z\sim0.5$--$5$ and peaking
near $z\simeq2$. Over the multipoles relevant to current CMB-lensing
measurements, this projection samples quasi-linear and nonlinear scales, roughly $k\sim0.1$--$1\,h\,\mathrm{Mpc}^{-1}$, although the low-redshift and high-$L$ tails can extend to a few $h\,\mathrm{Mpc}^{-1}$.

On these scales, nonlinear gravitational collapse boosts the matter power
spectrum in CDM cosmologies. For ULA dark matter, however, quantum pressure
suppresses structure growth below the axion Jeans scale and modifies the
abundance and structure of halos. The nonlinear response can therefore differ
qualitatively from the \lcdm{} case, and must be modelled with a prescription
appropriate for mixed-dark-matter cosmologies.

Throughout this section, we compare the effects of different nonlinear models on the matter and CMB power spectra. Our fiducial \Planck{} 2018 \lcdm{} baseline has parameters: Hubble constant $H_0 = 67.32 \, {\rm km} \, {\rm s}^{-1} \, {\rm Mpc}^{-1}$, baryon density $\Omega_{\rm{b}} h^2 = 0.022383$, cold dark matter density $\Omega_{\rm{c}} h^2 = 0.12011$, scalar spectral index $n_{\rm{s}} = 0.96605$, primordial amplitude $A_{\rm{s}} = 2.101 \times 10^{-9}$, optical depth $\tau = 0.0543$, and sum of the neutrino masses $\sum m_\nu = 0.06\,$eV.

\subsection{HMcode}

The Boltzmann code \texttt{CAMB} calculates nonlinear corrections to the matter power spectrum using \texttt{HMcode-2020} \cite{Mead:2020vgs}. This halo model (HM) implementation accurately and quickly predicts the nonlinear matter power spectrum for CDM, and is calibrated to simulations. \texttt{HMcode-2020} improved on the previous version by adding models for BAO damping and baryonic feedback on the power spectrum. Assuming all matter is cold and contained within halos, the nonlinear matter power spectrum is the sum of two terms, the one-halo term $P^{1\rm{h}}$ and the two-halo term $P^{2\rm{h}}$
\begin{equation}
    P(k) = (P^{1\rm{h}}(k)^{\alpha} + P^{2\rm{h}}(k)^{\alpha})^{1/\alpha} \,.
    \label{eq:cold_matter_power_spectrum}
\end{equation}
$P^{1\rm{h}}$ is the correlation within a single halo, given by an integral over the halo mass function (HMF) weighted by the squared Fourier transform of the Navarro-Frenk-White (NFW) halo density profile. It dominates on small scales. $P^{2\rm{h}}$ is the correlation between two different halos, and $\alpha$ is a parameter which smooths the transition between the two terms: $\alpha < 1$ gives a smoother transition, $\alpha > 1$ a steeper one. The value of $\alpha$ depends on the effective spectral index, $n_{\rm{eff}}$, which measures the slope of $\sigma(R)$, the RMS of the linear density field smoothed on scale $R$, evaluated at the nonlinear scale $R_{\rm{nl}}$ (defined by $\sigma^{2}(R_{\rm{nl}}) = 1$).

Internally, \texttt{HMcode-2020} constructs a ``cold'' power spectrum (composed of CDM and baryons) from the total matter power spectrum, $P(k)$. This treatment originates from cosmologies with massive neutrinos, where the nonlinear spectrum is computed for the non-neutrino (cold) component. This cold spectrum is then used to compute $\sigma(R)$, the nonlinear scale $R_{\rm{nl}}$, the effective spectral index $n_{\rm{eff}}$, and other halo model parameters. Crucially, when a ULA component is included
its perturbations are incorporated in the total matter transfer function. As a result, the axion-induced small-scale suppression contaminates the cold spectrum that \texttt{HMcode} computes internally. This simultaneously shifts several \texttt{HMcode} internal parameters: the effective small-scale $\sigma(R)$ is suppressed, $n_{\rm eff}$ becomes more negative, the inferred nonlinear scale $R_{\rm{nl}}$ becomes smaller and the smoothing parameter $\alpha$ decreases to its numerical safeguard floor of 0.5. The floor is calculated from the lower bound on the effective spectral index, $n_{\rm{eff}} > -3$.

\texttt{HMcode-2020} has been calibrated using $\Lambda$CDM simulations, and includes a non-cold matter component assumed to behave like massive neutrinos. However, it has not been calibrated to cosmologies containing ULAs. The code therefore takes the total linear matter power spectrum $P(k)$ --- which already includes the axion-induced small-scale suppression --- as input, and computes the nonlinear power spectrum as if only the CDM structure formation were reduced on those scales. This is an incorrect treatment as the halo mass function, concentration-mass relation, and virial properties all depend on whether the dark matter is cold or has a de Broglie-scale cutoff. Additionally, ULAs differ from massive neutrinos, which suppress small-scale structure formation below their free-streaming scale because of their thermal velocities. While these velocities inhibit neutrino clustering on small scales, the low-velocity tail of the neutrino distribution can still cluster within deep CDM halo potential wells. By contrast, ULA suppression arises from wave pressure associated with the characteristic axion Jeans scale rather than particle free-streaming \cite{Lesgourgues:2006nd, Massara:2014kba}. 

Within \texttt{AxiCAMB} we modify \texttt{HMcode-2020} to use the correct background expansion history, including the axion contribution, in place of the default $\Lambda$CDM background. However, nonlinear modelling continues to assume all dark matter is cold. Consequently, the HMF, concentration-mass relation and $\sigma(R)$ are computed from the total matter power spectrum, but do not distinguish between the cold and axion components. We refer to this model as the ``naive'' nonlinear model throughout this work.

\begin{figure*}
    \includegraphics[width=\textwidth]{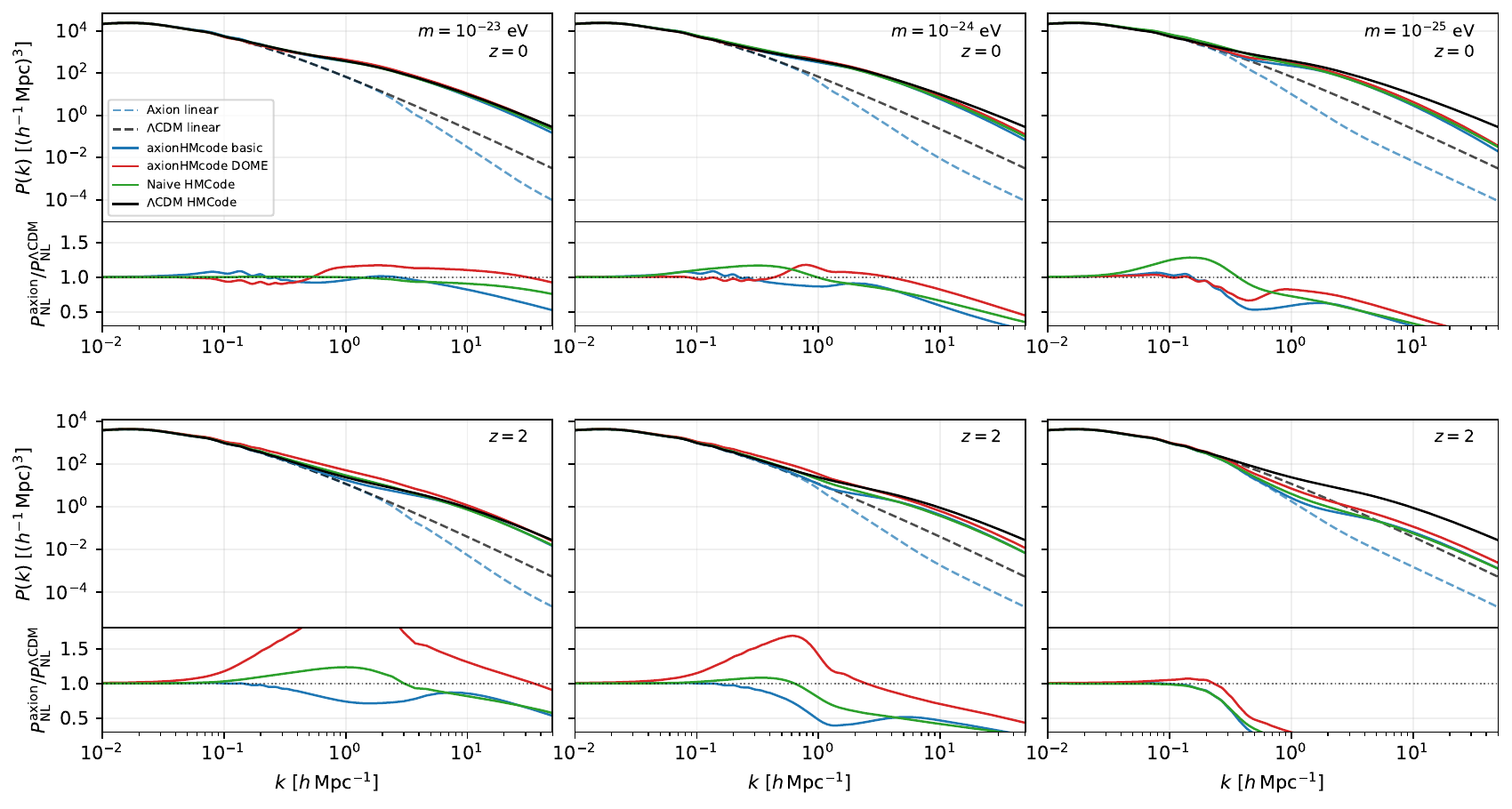}
\caption{The matter power spectra as a function of wave number $P(k)$, for $f_{\rm ax} = 0.3$ and three axion masses: $m = 10^{-23}\,$eV (left), $10^{-24}\,$eV (centre), and $10^{-25}\,$eV (right), at $z = 0$ (top panels) and $z = 2$ (bottom panels). The upper sub-panels show the matter power spectra: dashed lines are linear spectra (blue: axion, grey: $\Lambda$CDM), solid lines are nonlinear (blue: \texttt{axionHMcode} basic, red: \texttt{axionHMcode} DOME, green: naive \texttt{HMcode}, black: $\Lambda$CDM with \texttt{HMcode-2020}). The lower sub-panels show the ratio of the axion nonlinear matter power spectra to $\Lambda$CDM with \texttt{HMcode-2020}.}
    \label{fig:pk_masses}
\end{figure*}

\Cref{fig:pk_masses} shows the ratio of the axion to $\Lambda$CDM nonlinear matter power spectra for axion masses $m = 10^{-25}$, $10^{-24}$, and $10^{-23}\,$eV, for $f_{\rm{ax}}=0.3$, at redshifts $z = 0$ and $z = 2$. The upper panels show the absolute spectra and lower panels the ratio of the axion nonlinear spectrum to the $\Lambda$CDM nonlinear spectrum computed with \texttt{HMcode-2020}. In each panel, we compare the different nonlinear models. 

The axion Jeans scale, $k_{\rm J}$, marks the approximate transition between large-scale CDM-like behaviour and the small-scale suppression caused by oscillations. The Jeans scale is related to the axion mass, $k_{\rm{J}} \propto m^{1/2}$, and as such the onset of the small-scale suppression is shifted to higher wavenumbers for heavier masses. This behaviour can be seen in Fig.~\ref{fig:pk_masses}. At $z=0$, the $m=10^{-23}\,\rm{eV}$ case is nearly indistinguishable from $\Lambda$CDM in the naive nonlinear $P(k)$, with less than 1\% differences at scales $k \lesssim 1\, h\,\rm{Mpc}^{-1}$. However, at $z=2$, there is a spurious excess of approximately (15--25)\% from naive \texttt{HMcode} at scales $k \sim (0.3$--$1) \, h\,\rm{Mpc}^{-1}$. Although the linear suppression is still weak on these scales at $z=2$, it shifts \texttt{HMcode}'s internal parameters --- in particular $n_{\rm{eff}}$, $R_{\rm{nl}}$ and $\alpha$ --- and the resulting smoothing of the one-to-two transition produces enhanced power in the quasi-linear regime. As the CMB lensing kernel peaks at $z\sim2$ \cite{Lewis:2006fu, Hassani:2015zat}, the $z=2$ behaviour is more relevant than the $z=0$ behaviour for the lensed CMB power spectra.

For $m = 10^{-25}\,\rm{eV}$, the axion cutoff enters at larger scales, where linear suppression is already significant. Therefore, the naive nonlinear correction does not produce a cleanly isolated quasi-linear excess. In contrast, for $m = 10^{-24}\,\rm{eV}$, the cutoff falls inside the quasi-linear regime, where shifts in \texttt{HMcode}'s internal parameters produce the largest spurious enhancement relative to a proper nonlinear treatment. Consequently, $m = 10^{-24}\,\rm{eV}$ is the mass most sensitive to the choice of nonlinear prescription, motivating our focus on the mass range $10^{-25} \leq m/\rm{eV} \leq 10^{-23}$.

This spurious enhancement to the nonlinear matter power spectrum can be explained in terms of the smoothing parameter $\alpha$. In the quasi-linear regime, where $P_{\rm{1h}} \approx P_{\rm{2h}} \equiv P_0$, the nonlinear matter power spectrum becomes $P_{\rm{nl}} \approx 2^{1/\alpha} P_{0}$. Therefore, a small $\alpha$ can increase the overall power spectrum. For example, $\alpha = 1$ gives $P_{\rm{nl}} = 2P_0$; $\alpha = 0.7$ gives $P_{\rm{nl}} \approx 2.7P_0$; and $\alpha = 0.5$ gives $P_{\rm{nl}} = 4P_0$. As a result, although the individual halo terms are both suppressed compared to $\Lambda$CDM, the total power spectrum is enhanced. This excess is an artefact of the smoothing interpolation.

\subsection{axionHMcode}

\texttt{axionHMcode} \cite{Vogt:2022bwy} introduces a halo
model for mixed dark matter cosmologies containing a subdominant ULA
component. In this prescription, ULAs cluster in the potential wells of
cold halos on scales larger than the axion Jeans scale, while a cut-off halo
mass accounts for the suppression of ULA clustering below the Jeans scale.
The non-axion clustered component contains both CDM and baryons.

Since not all of the axion component is bound in halos, the standard
\texttt{HMcode} matter power spectrum is generalised to include cold,
cross, and axion contributions:
\begin{equation}
\begin{split}
    P(k) ={}&
    \left(\frac{\Omega_{\rm cb}}{\Omega_{\rm m}}\right)^2
    P_{\rm cb}(k)
    + \frac{2\Omega_{\rm cb}\Omega_{\rm ax}}{\Omega_{\rm m}^{2}}
    P_{\rm cb,ax}(k)  \\
    &+
    \left(\frac{\Omega_{\rm ax}}{\Omega_{\rm m}}\right)^2
    P_{\rm ax}(k) .
\end{split}
\label{eq:axionhmcode_total_power}
\end{equation}
Here $\Omega_{\rm cb}\equiv\Omega_{\rm cdm}+\Omega_{\rm b}$ is the
CDM-plus-baryon density, $\Omega_{\rm ax}$ is the axion density, and
$\Omega_{\rm m}=\Omega_{\rm cb}+\Omega_{\rm ax}$. The spectrum
$P_{\rm cb}(k)$ is the nonlinear CDM-plus-baryon power spectrum, computed
using the standard halo-model prescription as in
Eq.~\ref{eq:cold_matter_power_spectrum}. The terms $P_{\rm cb,ax}(k)$ and
$P_{\rm ax}(k)$ are the corresponding cold--axion cross spectrum and axion
auto-spectrum, respectively.

Unlike the cold component, the ULA component is not assumed to be entirely
halo-bound. Instead, the ULA overdensity is split into a clustered,
halo-bound contribution and an unclustered linear contribution. This allows
the cross and axion spectra to account for the absence of ULA clustering on
scales below the Jeans scale. The axion dependence enters through the density
weighting in Eq.~\eqref{eq:axionhmcode_total_power}, the clustered fraction
and cut-off associated with the axion Jeans scale, and calibrated
modifications to the halo-model ingredients.

There are two versions of \texttt{axionHMcode}. The first version, referred
to here as the ``basic'' prescription \cite{Vogt:2022bwy}, was calibrated
against ULA simulations for axion masses in the range
$10^{-33} \leq m/\mathrm{eV} \leq 10^{-21}$ and axion fractions
$f_{\rm ax}<0.5$. The latter is a restriction of the biased-tracer approach.
In this version, the one-halo and two-halo terms are summed without
additional smoothing, corresponding to $\alpha=1$ in
Eq.~\ref{eq:cold_matter_power_spectrum}. The two contributions are therefore
treated independently.

The second version, referred to here as the ``DOME'' prescription
\cite{Dome:2024hzq}, was calibrated against a broader suite of mixed dark
matter simulations with improved modelling of the concentration--mass
relation. Its calibration applies to axion fractions $f_{\rm ax}\leq0.3$
and axion masses close to $m=10^{-24.5}\,\mathrm{eV}$. DOME modifies the
transition between the one-halo and two-halo terms by replacing the single
\texttt{HMcode} smoothing parameter with independent smoothing parameters
for the cold--cold and cold--axion spectra, calibrated to simulations.

In the DOME version, the calibrated smoothing parameters can be driven
below the naive \texttt{HMcode} floor of $\alpha=0.5$ at
$z\sim2$ for $m\sim10^{-24}\,\mathrm{eV}$. This broadens the transition
between the one-halo and two-halo terms, producing a quasi-linear boost.
We verify that this is primarily a smoothing effect: enabling DOME-style
smoothing in the basic version reproduces the quasi-linear boost. This
enhancement has the same mathematical origin as the naive
\texttt{HMcode} excess, namely a low value of the smoothing parameter, but
its interpretation is different. In DOME it is a simulation-calibrated
feature of the mixed-dark-matter halo model, whereas in the naive case it
arises from applying a CDM-calibrated nonlinear prescription to a
mixed-dark-matter power spectrum outside its intended regime of validity.

\subsection{Integration in AxiCAMB}

Within \texttt{AxiCAMB}, we have implemented a new nonlinear interface which allows for the direct integration of external nonlinear models into the Boltzmann solver. At each likelihood evaluation, the following steps are taken. First, \texttt{AxiCAMB} computes the linear cold matter power spectrum and linear total matter power spectrum. Second, \texttt{axionHMcode} takes these power spectra together with the axion mass and physical density as input, and computes the nonlinear power spectrum. The nonlinear-to-linear power spectrum ratio is then calculated and passed back to \texttt{AxiCAMB}, after which the lensed angular power spectrum, $C_{\ell}$, is computed using the correct nonlinear power spectrum. This approach is efficient: the Boltzmann equations are solved once per evaluation, with the external nonlinear ratio applied directly to the linear results. The nonlinear corrections remain self-consistent with the sampled cosmology at each MCMC step. The updated \texttt{AxiCAMB} code is publicly available, allowing users to integrate their own nonlinear prescriptions. See Fig.~\ref{fig:pk_masses} for the resulting matter power spectrum comparison.

\subsection{Impact on CMB lensing}

\begin{figure*}
    \includegraphics[width=\textwidth]{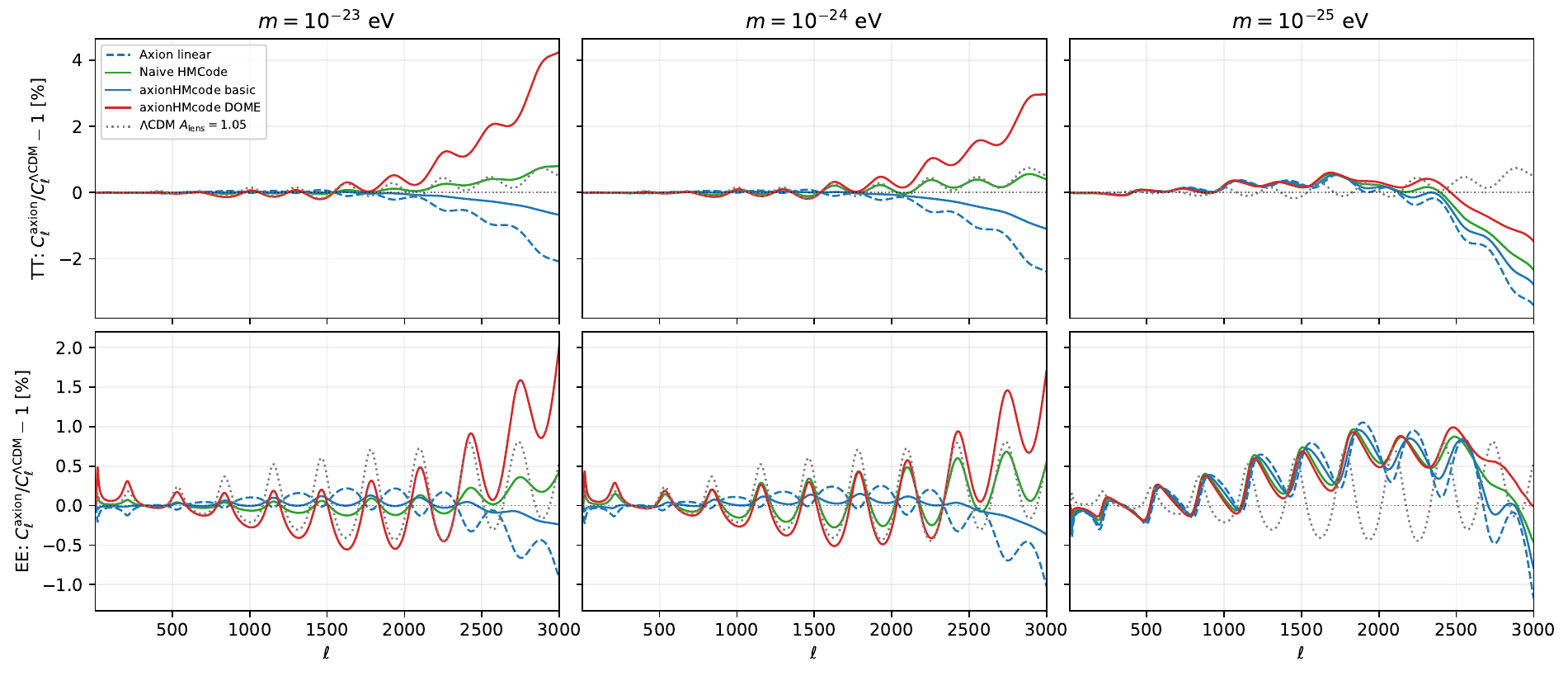}
\caption{The percentage difference in the lensed 
CMB TT (upper) and EE (lower) angular power spectrum relative to $\Lambda$CDM, $C_{\ell}^{\rm ax} /C_{\ell}^{\Lambda{\rm CDM}} -1$, for $f_{\rm ax} = 0.3$ and axion masses: $m = 10^{-23}\,$eV (left), $10^{-24}\,$eV (centre), and $10^{-25}\,$eV (right). The dashed blue lines show the axion linear spectra. Nonlinear spectra are shown for $\Lambda$CDM with $A_{\rm lens} =1.05$ (grey dotted), naive HMcode (green), axionHMcode basic (blue), and axionHMcode DOME (red).}
    \label{fig:cls}
\end{figure*}

Differences in the nonlinear matter power spectrum $P(k)$ propagate to the lensed CMB through the lensing potential $\phi$. \Cref{fig:cls} shows the fractional difference from $\Lambda$CDM for the axion lensed $C_{\ell}^{TT}$ and $C_{\ell}^{EE}$ CMB power spectra. With no nonlinear corrections, the reduced lensing power leads to a suppression of the temperature spectrum, reaching approximately $-2\%$ at multipoles $\ell \sim 3000$, independent of axion mass. With the naive \texttt{HMcode} prescription, the suppression is partially compensated, reducing differences to roughly $\pm 0.5\%$ for $m=10^{-23}$ and $10^{-24}\,$eV. However, a large suppression remains for $10^{-25}\,$eV. The \texttt{axionHMcode} basic version follows the linear result more closely, and preserves the lensing suppression, with an approximately $- 1\%$ change in TT at $\ell \sim 3000$. In contrast, the DOME version produces a large nonlinear boost, with oscillatory residuals up to $\sim 3\%$ in TT at $\ell \sim 3000$. Thus the choice of nonlinear model leads to a spread of several percent at $\ell \gtrsim 2000$ in predicted CMB spectra. In comparison, the $\Lambda$CDM model with lensing potential power spectrum rescaling parameter $A_{\rm{lens}} = 1.05$ produces approximately $0.5\%$ oscillatory differences. The uncertainty from axion nonlinear modelling is therefore much larger. 

As shown in Fig.~\ref{fig:cls}, the lensed CMB spectra for the axion model with naive \texttt{HMcode} nonlinear corrections closely resemble the $\Lambda$CDM model with $A_{\rm{lens}} > 1$ for $m= 10^{-23}$ and $10^{-24}\,$eV. This is because the naive model over-predicts the nonlinear boost, which in turn produces excess lensing power. As a result, an axion signal under the naive prescription can be partially mimicked or absorbed by the well-known preference for $A_{\rm{lens}} > 1$ in \Planck{} data \cite{Planck:2018vyg}. In contrast, when the nonlinear suppression of power expected in MDM cosmologies is modelled more accurately, the lensing excess is reduced. This weakens the apparent preference for an axion DM component.

\begin{figure}
    \includegraphics[width=\columnwidth]{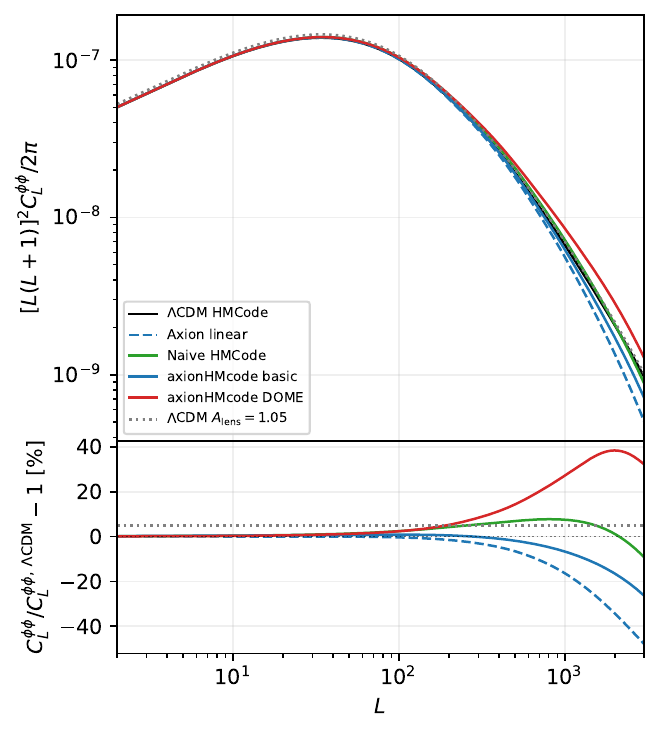}
\caption{The CMB lensing potential power spectrum, $C_L^{\phi\phi}$, for $m = 10^{-24}\,$eV and $f_{\rm{ax}}=0.3$. The upper sub-panel shows absolute spectra, and the lower sub-panel shows the percentage difference from $\Lambda$CDM with \texttt{HMcode-2020}. The line colours and styles are the same as in previous figures.}
    \label{fig:clpp}
\end{figure}

\Cref{fig:clpp} shows the CMB lensing potential power spectrum, $C_{L}^{\phi\phi}$, and its percentage difference from the $\Lambda$CDM spectrum calculated with \texttt{HMcode}. We see that the lensing potential is significantly more sensitive to nonlinear model choice than the primary CMB. In particular, differences of approximately $(5$--$15)\%$ at lensing multipoles $L \gtrsim 500$ are observed between the naive and \texttt{axionHMcode} models, compared to the sub-percent differences seen in the lensed $C_{\ell}^{TT}$ and $C_{\ell}^{EE}$ power spectra. Therefore, analyses incorporating measurements of the lensing potential spectrum $C_{L}^{\phi\phi}$ (such as from ACT, \Planck{}, Simons Observatory and other future experiments) will be strongly affected by the choice of nonlinear model.

\section{Data and Method} \label{sec:method}

\subsection{Data}

We use the public \Planck{} 2018 low-$\ell$ temperature (TT) and SRoll2 polarisation (EE) likelihoods \cite{Delouis:2019bub}. SRoll2 includes an improved mapmaking algorithm to correct for instrumental effects which affected polarised large-scale data. This reduced the variance at $\ell < 6$ by a factor $\sim2$ with respect to previous data, thus improving parameter constraints sensitive to the reionisation peak at low multipoles (such as the optical depth $\tau$). We combine these with the foreground-marginalised ACT DR6 CMB-only TT/TE/EE likelihood (\texttt{ACT-lite}), and a cut version of the high-$\ell$ \Planck{} likelihood, restricted to $\ell < 1000$ in TT and $\ell < 600$ in TE/EE \cite{AtacamaCosmologyTelescope:2025nti}, to minimise overlap between the two experiments. 

We also use the DESI DR2 BAO likelihood \cite{DESI:2025zgx} using all galaxy and quasar tracers. This release uses three years of data and comprises BAO measurements from more than 14 million galaxies and quasars, with statistical precision improved by $(30$--$50)\%$ over DESI DR1. 

To complement the primary CMB power spectra, we use ACT DR6 CMB lensing data \cite{ACT:2023dou, ACT:2023kun}. This release provides a $43\sigma$ measurement of the lensing power spectrum across multipoles $40 < L < 763$. We combine this with the \Planck{} PR4 NPIPE lensing spectrum, which provides a $42\sigma$ measurement across multipoles $8 < L < 400$ \cite{Carron:2022eyg}. The effective signal-to-noise of this combination is $58\sigma$, providing a precise measurement of the CMB lensing power spectrum. 

We consider two data combinations for our MCMC analyses. In the first (denoted \pact{}), only primary CMB spectra (\Planck{} + ACT) are used, with lensing entering through its effect on $C_\ell^{TT}$ and $C_\ell^{EE}$. The second (denoted \pactlb{}) additionally uses the combined \Planck{} \cite{Carron:2022eyg} + ACT \cite{ACT:2023kun,ACT:2023dou} likelihood for the CMB lensing potential power spectrum $C_L^{\phi\phi}$, and DESI DR2 BAO measurements. As shown in Fig.~\ref{fig:clpp}, $C_{L}^{\phi\phi}$ is more sensitive to the choice of nonlinear model than the primary CMB. Therefore, comparing these two combinations directly quantifies the impact of nonlinear modelling on the constraints.

\subsection{MCMC setup}

We perform a MCMC analysis using the \textsc{Cobaya}~\citep{Torrado:2020dgo} sampler, with \texttt{AxiCAMB} as the Boltzmann solver. We assume a spatially flat cosmology and adopt flat priors for the six base cosmological parameters:
\[
  \{\,H_0,\, \Omega_\mathrm{D} h^2,\, \Omega_\mathrm{b} h^2,\, n_\mathrm{s},\, \log(10^{10} A_\mathrm{s}),\, \tau\},
\]
where $H_0$ is the Hubble constant, $\Omega_\mathrm{D} h^2$ is the total dark matter physical density (cold plus axion), $\Omega_\mathrm{b} h^2$ is the baryon physical density, $n_\mathrm{s}$ is the scalar spectral index, $A_\mathrm{s}$ is the primordial amplitude, and $\tau$ is the optical depth. Following Ref.~\cite{Rogers:2023ezo}, we perform separate MCMC runs at three fixed axion masses, $m\in\{10^{-25},10^{-24},10^{-23}\}\,\mathrm{eV}$, rather than sampling the mass jointly with the other cosmological parameters.

With the \pact\, data combination, for each mass we run chains for five nonlinear model choices:
\begin{enumerate} [noitemsep,nolistsep] 
    \item naive \texttt{HMcode-2020},
    \item \texttt{axionHMcode} basic,
    \item \texttt{axionHMcode} DOME,
    \item \texttt{axionHMcode} basic with nuisance parameter marginalisation, 
    \item \texttt{axionHMcode} DOME with nuisance parameter marginalisation.
\end{enumerate}
With the \pactlb\, data combination, we run chains for choices 1--3, without nuisance parameter marginalisation. This is because marginalising over these parameters in the primary-CMB-only runs had negligible impact on the results.  
For naive runs, we use a flat prior for the axion fraction,
\begin{equation}
    0 \leq f_{\rm{ax}} \leq 1.
\end{equation}
For basic runs the flat prior is restricted to $0 \leq f_{\rm{ax}} \leq 0.5$ and for DOME runs to $ 0 \leq f_{\rm{ax}} \leq 0.3$. This keeps the axion fraction within the HM limit and calibration limits respectively. 

 The DOME calibration is centred on axion masses close to
$m=10^{-24.5}\,\mathrm{eV}$.
Our fixed-mass analyses at $m=10^{-25}$ and
$10^{-24}\,\mathrm{eV}$ lie half a decade below and above this calibration
scale, respectively, while the $m=10^{-23}\,\mathrm{eV}$ case lies
1.5 decades above it. The DOME results should therefore be interpreted as
sensitivity tests of the calibrated fitting formula, with the
$m=10^{-23}\,\mathrm{eV}$ case representing the largest extrapolation in
mass.

Following Ref.~\cite{Dentler:2021zij}, we adopt flat priors on the \texttt{axionHMcode} nuisance parameters 
\begin{align}
    \alpha_{1} &: [0.6, 2.0], \nonumber \\
    \alpha_{2} &: [1.43, 2.54], \nonumber \\
    \gamma_{1} &: [5.0, 45.0], \nonumber \\
    \gamma_{2} &: [-0.37, -0.23], 
\end{align}
 where $\alpha_{1,2}$ control the suppression of the HMF. $\alpha_{1}$ sets how sharply the HMF transitions from agreement with CDM to suppression, while $\alpha_{2}$ controls the steepness of this suppression. $\gamma_{1,2}$ control the additional suppression of the halo concentration parameter (which quantifies how centrally concentrated a halo is) seen in simulations: $\gamma_{1}$ shifts the position of the cut-off scale, and $\gamma_{2}$ controls its slope. 

We use a switch $m/H_{*}=50$ and \texttt{accuracy=2.5} to balance speed and precision, ensuring the CMB power spectra are accurate across all masses and nonlinear models. Additionally, we choose \texttt{AxiCAMB} accuracy settings within \texttt{Cobaya} to match the ACT DR6 axion MCMC calculations \cite{AtacamaCosmologyTelescope:2025nti}.

For each run we use the Metropolis-Hastings sampler with convergence criterion $R-1 \le 0.02$. Once complete, we discard a burn-in portion of each chain and use the final posterior distribution to infer cosmological and axion parameters. After convergence, we refine the maximum-likelihood estimate using BOBYQA minimisation \cite{cartis2019improving, cartis2022escaping}, with the chain best-fit point as the starting point.

\section{Results} \label{sec:results}

\subsection{Fixed-parameter scan}

\begin{figure*}
    \includegraphics[width=\textwidth]{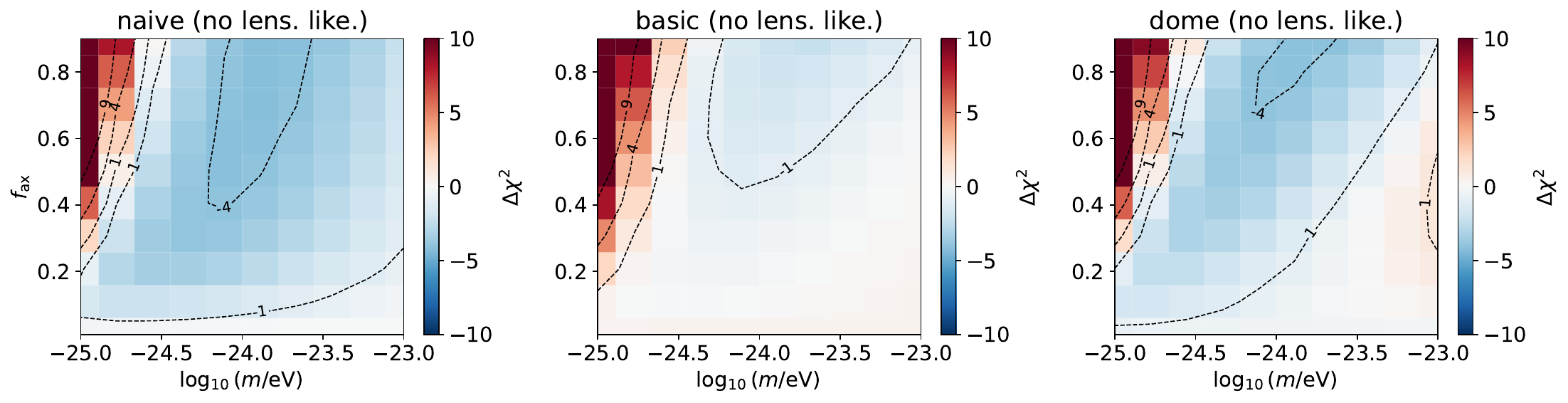}
\caption{The $\Delta\chi^2$ of the naive \texttt{HMcode} and \texttt{axionHMcode} basic and DOME versions relative to the best-fit $\Lambda$CDM model from \Planck{} as a function of $(\log_{10}( m), f_{\rm ax})$, excluding the lensing likelihood. Contours show $\Delta\chi^2 = \pm 1, \pm 4, \pm 9$, corresponding to the nominal 1, 2, and $3\sigma$ thresholds for one parameter.}
    \label{fig:chi2_nolenslike}
\end{figure*}

\begin{figure*}
    \includegraphics[width=\textwidth]{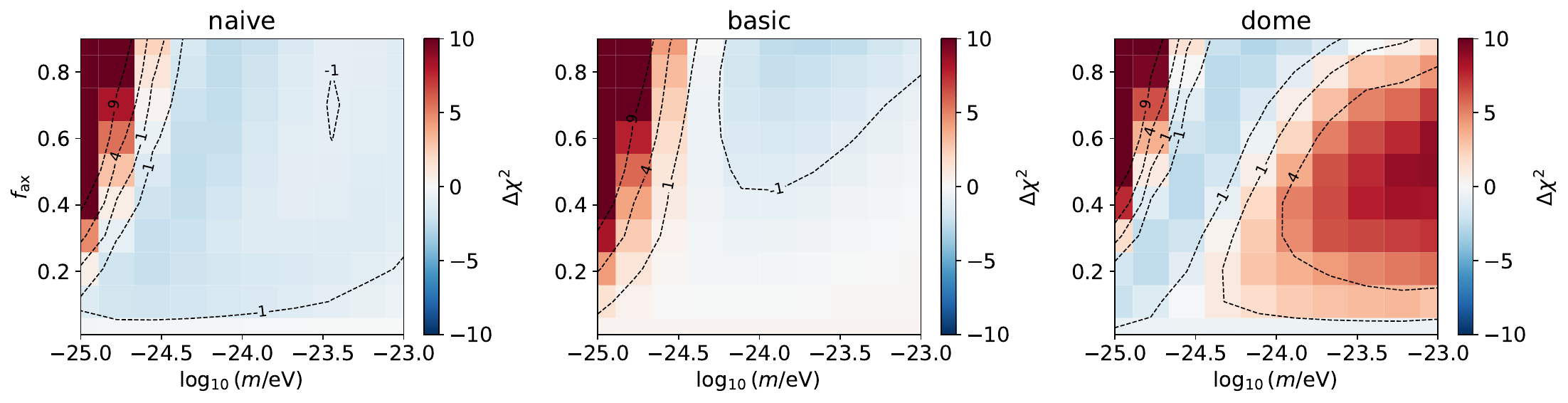}
\caption{Same as Fig.~\ref{fig:chi2_nolenslike}, but including the lensing likelihood.}
    \label{fig:chi2_naive}
\end{figure*}

Before performing the MCMC runs, we conduct a diagnostic scan over the
axion mass and fraction, $(m,f_{\rm ax})$, with all other cosmological
parameters fixed to the \lcdm{} best fit. The purpose of this scan is not
to derive parameter constraints, but to identify the regions of parameter
space where the inferred goodness of fit is most sensitive to the nonlinear
modelling prescription. Figures~\ref{fig:chi2_nolenslike}
and~\ref{fig:chi2_naive} show the resulting $\Delta\chi^2$ relative to
\lcdm{}, excluding and including the lensing likelihood respectively.
We define
\begin{equation}
    \Delta\chi^2=\chi^2_\mathrm{ax}-\chi^2_{\Lambda{\rm CDM}},
\end{equation}
where negative values indicate that the axion model gives a better fit than
\lcdm{} for the fixed cosmological parameters used in the scan. The effective
$\chi^2$ is defined as
\begin{equation}
    \chi^{2}_{\rm eff} = \chi^{2} - 2\ln p_{\rm cal},
\end{equation}
where $p_{\rm cal}$ denotes the calibration prior included in the likelihood.

At each fixed axion mass, the scan varies only one additional parameter,
$f_{\rm ax}$. We therefore show contours corresponding to
$|\Delta\chi^2|=1,4,$ and $9$ as nominal $1,2,$ and $3\sigma$ guides. These should not be interpreted as formal significances: the likelihood
need not be Gaussian, $f_{\rm ax}=0$ lies on a prior boundary, and the remaining cosmological parameters are held fixed.
The MCMC analysis below is therefore required to obtain marginalised
constraints and robust goodness-of-fit comparisons.

For this diagnostic scan we also show the behaviour of the
\texttt{axionHMcode} prescriptions outside their formal calibration ranges.
This is done deliberately in order to assess how the $\Delta\chi^2$ landscape
responds to the nonlinear prescription over the same $(m,f_{\rm ax})$ grid
used for the naive model. Points outside the calibrated domain should be
viewed only as extrapolations of the fitting formulae, not as validated
predictions or constraints.

As shown in Fig.~\ref{fig:chi2_nolenslike}, the naive nonlinear model
produces a region with $\Delta\chi^2 \lesssim -4$ for
$f_{\rm ax}\gtrsim 0.2$ and $m\gtrsim 10^{-24.5}\,\mathrm{eV}$, corresponding
to a nominal $>2\sigma$ improvement.
This apparent improvement is strongly model-dependent. With the
\texttt{axionHMcode} basic prescription, the magnitude of the improvement is
reduced to the nominal $1\sigma$ level, while the DOME prescription gives a
different trend, particularly at masses $m\gtrsim10^{-23.5}\,\mathrm{eV}$.

Including the lensing likelihood, Fig.~\ref{fig:chi2_naive}, further reduces
the apparent improvement in the naive model, with the region near
$m\sim10^{-23.5}\,\mathrm{eV}$ falling to roughly the nominal $1\sigma$ level.
For the DOME prescription, the same region instead gives positive
$\Delta\chi^2$, with \lcdm{} preferred at the level of
$\Delta\chi^2\gtrsim4$ for $m\gtrsim10^{-24}\,\mathrm{eV}$. Overall, the
fixed-parameter scan shows that the inferred $\Delta\chi^2$ varies
substantially between nonlinear prescriptions across the mass range of
interest. This motivates the three fixed masses used in the MCMC analysis,
$m=10^{-25}$, $10^{-24}$, and $10^{-23}\,\mathrm{eV}$, which sample the
regime where nonlinear modelling has the largest impact on the cosmological
fit. While the grid scan is indicative, the analysis presented below is required to capture degeneracies with the other cosmological parameters and to marginalise over nonlinear nuisance parameters.

\subsection{MCMC}

\begin{figure*}
    \includegraphics[width=\textwidth]{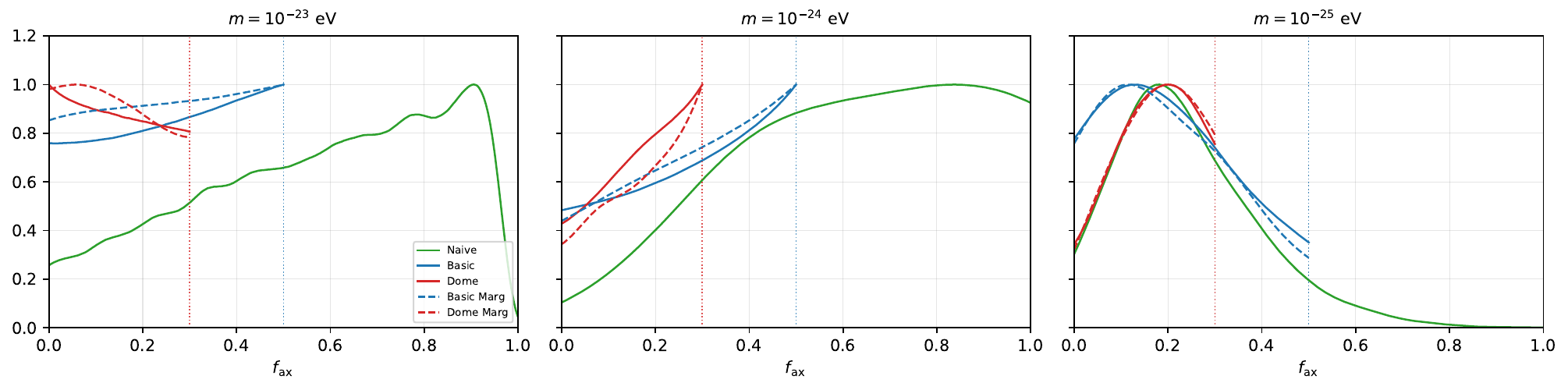}
\caption{One-dimensional posterior distributions for $f_{\rm ax}$
 using the P-ACT data combination for axion masses: $10^{-23}\,$eV (left), $10^{-24}\,$eV (centre) and $10^{-25}\,$eV (right). Results are shown for naive \texttt{HMcode} (green), \texttt{axionHMcode} basic (blue), and \texttt{axionHMcode} DOME (red). Dashed curves include marginalisation over the \texttt{axionHMcode} nuisance parameters. Vertical dotted lines indicate the upper prior limits for the basic and DOME models.}
    \label{fig:1d_marginals}
\end{figure*}

\begin{figure*}
    \includegraphics[width=\textwidth]{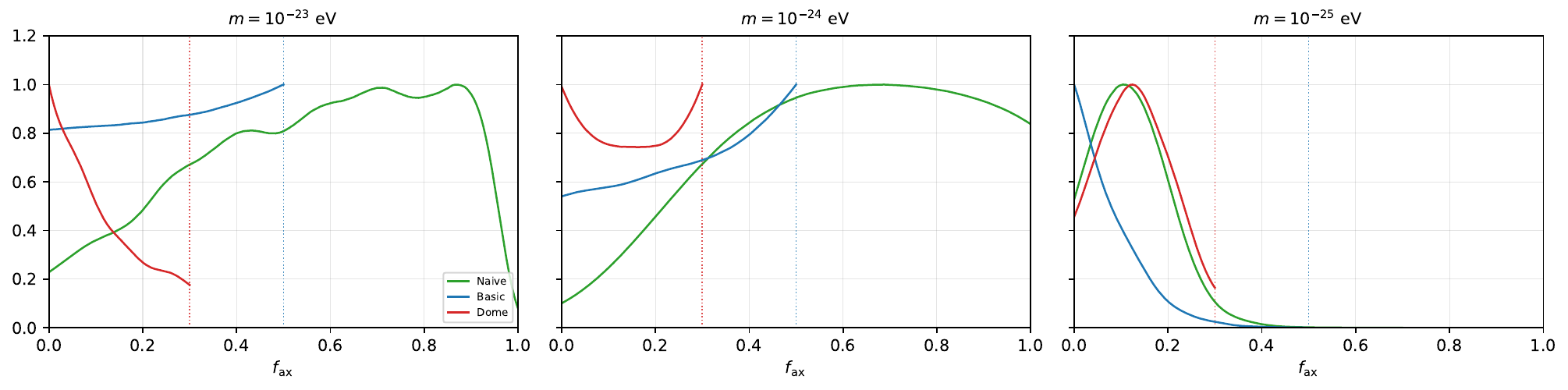}
\caption{Same as Fig.~\ref{fig:1d_marginals}, but for the \pactlb\, data combination and without \texttt{axionHMcode} nuisance parameter marginalisation.}
    \label{fig:1d_marginals_lens}
\end{figure*}

We now present the results of our analysis comparing the different nonlinear treatments at masses $m = 10^{-25}$, $10^{-24}$, and $10^{-23}\,$eV, where nonlinear modelling has the largest impact. \Cref{tab:chi2_f_ax} summarises the minimum effective $\chi^{2}$ for each model configuration along with the $\Delta\chi^{2}$ relative to $\Lambda$CDM, and shows the $95\%$ upper or lower limits, or posterior mean with $95\%$ credible interval, on the axion fraction $f_{\rm{ax}}$ and physical density $\Omega_{\rm{ax}}h^{2}$.

\begin{table}[]
\begin{tabular}{|l|l|l|l|l|l|}
\hline
Mass/eV                     & Non-Linear & $\chi^{2}$ & $\Delta \chi^{2}$ & $f_{\rm{ax}}$ & $\Omega_{\rm{ax}}h^{2}$ \\ \hline
\multirow{5}{*}{\begin{tabular}[c]{@{}l@{}}$10^{-23}$ \\ (\pact)\end{tabular}}   & N    & 779.95 & $-2.22$ & $0.58^{+0.40}_{-0.48}$ & $0.069^{+0.047}_{-0.059}$ \\ \cline{2-6} 
                                    & B    & 781.37 & $-0.80$ & ---                    & $< 0.0571$ \\ \cline{2-6} 
                                    & D    & 781.04 & $-1.13$ & ---                    & $< 0.0336$ \\ \cline{2-6} 
                                    & B (M)   & 781.40 & $-0.77$ & ---                    & $0.030^{+0.027}_{-0.030}$ \\ \cline{2-6} 
                                    & D (M)   & 781.10 & $-1.07$ & ---                    & $< 0.0336$ \\ \hline
\multirow{3}{*}{\begin{tabular}[c]{@{}l@{}}$10^{-23}$ \\ (\pactlb)\end{tabular}} & N  & 813.49 & $-2.98$ & $0.56^{+0.41}_{-0.46}$ & $0.066^{+0.047}_{-0.055}$ \\ \cline{2-6} 
                                    & B  & 816.16 & $-0.31$ & ---                    & $0.030^{+0.026}_{-0.029}$ \\ \cline{2-6} 
                                    & D   & 814.13 & $-2.34$ & ---                    & $< 0.0311$ \\ \hline
\multirow{5}{*}{\begin{tabular}[c]{@{}l@{}}$10^{-24}$ \\ (\pact)\end{tabular}}   & N    & 777.61 & $-4.56$ & $> 0.174$              & $0.072^{+0.047}_{-0.053}$ \\ \cline{2-6} 
                                    & B    & 780.18 & $-1.99$ & ---                    & $< 0.0576$ \\ \cline{2-6} 
                                    & D    & 779.03 & $-3.14$ & ---                    & $< 0.0344$ \\ \cline{2-6} 
                                    & B (M)  & 780.15 & $-2.02$ & ---                    & $0.034^{+0.026}_{-0.030}$ \\ \cline{2-6} 
                                    & D (M)  & 779.08 & $-3.09$ & ---                    & $< 0.0345$ \\ \hline 
\multirow{3}{*}{\begin{tabular}[c]{@{}l@{}}$10^{-24}$ \\ (\pactlb)\end{tabular}} & N  & 811.14 & $-5.33$ & $> 0.169$              & $0.069^{+0.047}_{-0.052}$ \\ \cline{2-6} 
                                    & B  & 815.50 & $-0.97$ & ---                    & $0.032^{+0.026}_{-0.030}$ \\ \cline{2-6} 
                                    & D  & 814.03 & $-2.44$ & ---                    & $< 0.0337$ \\ \hline
\multirow{5}{*}{\begin{tabular}[c]{@{}l@{}}$10^{-25}$ \\ (\pact)\end{tabular}}   & N    & 779.38 & $-2.79$ & $< 0.533$              & $< 0.0639$ \\ \cline{2-6} 
                                    & B    & 781.08 & $-1.09$ & ---                    & $< 0.0540$ \\ \cline{2-6} 
                                    & D    & 778.74 & $-3.43$ & ---                    & $0.020^{+0.015}_{-0.017}$ \\ \cline{2-6} 
                                    & B (M)   & 781.13 & $-1.04$ & ---                    & $< 0.0534$ \\ \cline{2-6} 
                                    & D (M)   & 778.72 & $-3.45$ & ---                    & $0.020^{+0.015}_{-0.017}$ \\ \hline 
\multirow{3}{*}{\begin{tabular}[c]{@{}l@{}}$10^{-25}$ \\ (\pactlb)\end{tabular}} & N  & 815.27 & $-1.20$ & $< 0.267$              & $< 0.0314$ \\ \cline{2-6} 
                                    & B  & 816.53 & $+0.06$ & $< 0.219$              & $< 0.0258$ \\ \cline{2-6} 
                                    & D  & 812.68 & $-3.79$ & ---                    & $< 0.0303$ \\ \hline
\end{tabular}
\caption{The constraints on the axion fraction, $f_{\rm{ax}}$, and physical density, $\Omega_{\rm{ax}}h^{2}$, for $m=10^{-23}, 10^{-24}$ and $10^{-25}$ eV, as $95\%$ upper or lower limits, or posterior means with $95\%$ credible intervals. Dashes denote no constraint. Also shown are the minimum effective $\chi^{2}$ (including the calibration prior) and $\Delta\chi^{2}$ relative to $\Lambda$CDM for each nonlinear model and likelihood combination. The nonlinear models are naive (N), basic (B) and DOME (D). Marginalisation over \texttt{axionHMcode} parameters is denoted by M. Results with the lensing and BAO likelihoods are denoted \pactlb{}; all other runs include primary CMB spectra only and are denoted \pact{}. For the \pact\, and \pactlb\, likelihoods, the $\Lambda$CDM effective $\chi^{2}=782.17$ and $816.47$, respectively.}
\label{tab:chi2_f_ax}
\end{table}

To compute $\Delta\chi^{2}$ relative to $\Lambda$CDM, we run a separate \lcdm{} MCMC analysis for each data combination, with \texttt{HMcode-2020} for the nonlinear modelling. This gives effective minimised $\chi^{2} = 782.17$ for \pact{} and $\chi^{2} = 816.47$ for \pactlb{}. Note that for the basic and DOME prescriptions, the $f_{\rm{ax}} \to 0$ limit does not exactly reproduce our \lcdm{} reference; we have verified that the resulting $\Delta\chi^{2}$ offset is small ($\Delta\chi^{2} = -0.76$ and $-0.50$ for basic and DOME respectively).

Using the naive \texttt{HMcode-2020} nonlinear model, the axion model gives
an apparent improvement relative to \lcdm{} of $\Delta\chi^2=-4.56$ at
$m=10^{-24}\,\mathrm{eV}$. If this mass and nonlinear prescription were
treated as a pre-specified one-parameter comparison, this would correspond to
a local nominal significance of $\sim2.1\sigma$. However, no look-elsewhere
correction has been applied for the scan over axion mass, and the mapping
from $\Delta\chi^2$ to a Gaussian significance is only approximate because
$f_{\rm ax}=0$ is a prior boundary. We therefore interpret this as a mild,
model-dependent improvement in fit rather than evidence for an axion
component.
 
 For the \texttt{axionHMcode} basic version, this preference decreases to $\Delta \chi^2=-1.99$ (nominal $\sim 1.4\sigma$), while the DOME version gives an intermediate result. A similar trend is seen for $m=10^{-23}\,$eV. For $m=10^{-25}\,$eV, DOME \emph{increases} the preference compared to the naive model. These results indicate that, although there is a mild preference for axions as a fraction of DM, this preference is highly sensitive to the nonlinear prescription.

Marginalising over \texttt{axionHMcode} nuisance parameters produces only minor changes in $\Delta\chi^{2}$ --- small reductions for $m=10^{-25}\,$eV, slight increases at $m = 10^{-24}\,$eV, and mixed behaviour at $m=10^{-23}\,$eV. In all cases the changes are small, so the marginalisation has negligible impact on the apparent axion preference. The nuisance-parameter posterior also shows no degeneracy with the cosmological or axion-mass parameters.

Including the lensing likelihood adds direct sensitivity to the small-scale matter power spectrum, where the differences between nonlinear models are largest. As with the CMB-only data, the naive nonlinear modelling results in a large $\Delta\chi^{2}$ relative to \lcdm{} ($\Delta\chi^{2} \approx -2.98, -5.33, -1.20$ for $m = 10^{-23}, 10^{-24}, 10^{-25}\,$eV, respectively), while the basic version gives the smallest $|\Delta\chi^{2}|$ ($\Delta\chi^{2} \approx -0.31, -0.97, +0.06$) and the DOME version gives an intermediate reduction ($\Delta\chi^{2} \approx -2.34, -2.44, -3.79$). Including lensing data emphasises the sensitivity of the apparent axion signal to the choice of nonlinear model.

The choice of nonlinear model also affects the physical interpretation of the apparent axion signal.
\Planck{} 2018 data showed enhanced lensing smoothing, quantified by the parameter $A_{\rm{lens}}$, which artificially rescales the lensing power spectrum. \Planck{} found $A_{\rm{lens}} = 1.180 \pm 0.065$ \cite{Planck:2018vyg}. Combining with the ACT DR6 data reduces this to $A_{\rm{lens}} = 1.081 \pm 0.043$ \cite{AtacamaCosmologyTelescope:2025blo}, retaining the preference for lensing enhancement. When naive nonlinear modelling is used for ULAs, the resulting lensing enhancement is similar in magnitude to a \lcdm{} model with $A_{\rm{lens}} = 1.05$. The preference for $A_{\rm{lens}} > 1$ therefore means that an axion model with naive nonlinear modelling can improve the fit by producing a similar excess in lensing power. With more accurate nonlinear modelling, this spurious lensing excess is reduced, weakening the apparent preference for axions.

\Cref{fig:1d_marginals} shows the posteriors for the axion fraction, $f_{\rm{ax}}$, obtained from the \pact\, analysis for each nonlinear model. The upper limit on $f_{\rm{ax}}$ becomes stronger when \texttt{axionHMcode} nonlinear modelling is used. For $m=10^{-24}\,$eV, the basic version gives the tightest constraints, with DOME intermediate between basic and naive. When \texttt{axionHMcode} nuisance parameters are marginalised, the constraints remain close to the unmarginalised case for $m = 10^{-24}$ and $10^{-25}\,$eV, whereas for $m = 10^{-23}\,$eV the constraints weaken.

\Cref{fig:1d_marginals_lens} is the same as Fig.~\ref{fig:1d_marginals}, but for the data combination \pactlb. In this case, the constraints on $f_{\rm{ax}}$ are tighter for all axion masses, reflecting the lensing-data sensitivity to small-scale axion suppression. For $m = 10^{-25}\,$eV, the constraints strengthen significantly, with the basic version giving the tightest bound. For $m = 10^{-23}\,$eV, the DOME version also produces a much tighter constraint on $f_{\rm{ax}}$. 

Finally, we compare our analysis to the recent ACT DR6 axion result \cite{AtacamaCosmologyTelescope:2025nti}. ACT DR6 constrained the fraction of ULAs in total dark matter at a set of discrete masses in the range $10^{-28} < m/\rm{eV} < 10^{-24}$. They used the standard EFA implemented in \texttt{axionCAMB}, the basic \texttt{axionHMcode} version for nonlinear power spectra, and the \texttt{axionEMU} emulator to accelerate the computations. We compared their analysis (using the publicly available chains\footnote{\url{https://lambda.gsfc.nasa.gov/product/act/act_dr6.02/act_dr6.02_chains_prod_table.html}}) to our \texttt{AxiCAMB} MCMC results with the PH EFA, and find broad agreement. 

\section{Discussion and Conclusions} \label{sec:discussion}

We have performed a systematic study of how nonlinear modelling affects
ultralight axion constraints from primary CMB, CMB-lensing, and BAO data.
For axion masses in the range $m\sim10^{-25}$--$10^{-23}\,\mathrm{eV}$, the
axion Jeans scale overlaps with the quasi-linear scales that contribute
significantly to the CMB lensing potential. As a result, the inferred axion
fraction is sensitive not only to the linear suppression of structure, but
also to how this suppression is propagated into the nonlinear matter power
spectrum.

Our main result is that a naive CDM-calibrated nonlinear correction can
produce an apparent preference for a subdominant axion component. In
particular, at $m=10^{-24}\,\mathrm{eV}$ this prescription gives a nominal
$\sim2.1\sigma$ improvement relative to \lcdm{}. This should not be
interpreted as a robust statistical detection: the mapping from
$\Delta\chi^2$ to a Gaussian significance is only approximate, $f_{\rm ax}=0$
lies on a prior boundary, and the result is sensitive to the nonlinear model.
The preference is substantially reduced when using mixed-dark-matter
nonlinear prescriptions implemented in \texttt{axionHMcode}. This indicates
that the naive treatment can generate a spurious axion preference by
producing a lensing-like enhancement of the lensed CMB power spectra.

The size of this effect depends on the nonlinear prescription. The basic and
DOME calibrations of \texttt{axionHMcode} give different shifts in the
goodness of fit, with the DOME prescription generally lying between the
naive model and the basic calibration for the cases studied here. Across the
nonlinear models, the variation in the best-fit values is
$\Delta\chi^2\sim2$--$5$, comparable to or larger than the nominal statistical
preference for axions in the naive analysis. This makes nonlinear modelling a
leading systematic uncertainty in CMB-lensing constraints on ultralight
axions in this mass range. Where the \texttt{axionHMcode} prescriptions are
evaluated outside their formal calibration ranges, our results should be
viewed as diagnostic extrapolations rather than validated constraints.

The origin of the model dependence is clearest in the lensing potential power
spectrum. As shown in \cref{fig:clpp}, the different nonlinear prescriptions
produce changes of order $(5$--$15)\%$ in $C_L^{\phi\phi}$ at
$L\gtrsim500$, while the primary CMB spectra are less directly affected.
Since the lensed CMB spectra are sensitive to the integrated lensing
contribution, an inaccurate nonlinear correction can mimic an effective
increase in lensing power. This provides a concrete mechanism by which a
CDM-calibrated nonlinear prescription can bias axion constraints.

Several future directions are needed to reduce this uncertainty. First,
mixed-dark-matter nonlinear prescriptions should be further validated against
simulations spanning the axion masses, fractions, and redshifts relevant for
CMB lensing. Second, future analyses should propagate nonlinear-modelling
uncertainties into the parameter inference, rather than treating a single
nonlinear prescription as exact. Third, forecasts for upcoming high-precision
CMB-lensing measurements, such as those from Simons Observatory and related
experiments, would clarify how strongly nonlinear systematics will limit
future axion searches. Finally, joint analyses with galaxy surveys such as
DESI and Euclid will provide complementary information on the matter power
spectrum and may help distinguish a genuine axion-induced suppression from
modelling artefacts.

\section*{Acknowledgments}

AG and AM are supported by an STFC Consolidated Grant  [Grant Nos.  ST/X000672/1] and AG by a Leverhulme Research Fellowship [Grant No. RF-2025-282/4]. LG was supported by an STFC studentship [Grant No.\ ST/X508639/1]. For the purpose of open access, the authors have applied a CC BY public copyright license to any Author Accepted Manuscript version arising. We are grateful to Samuel Brieden, Catherine Heymans, Antony Lewis, David J. E. Marsh and Keir Rogers for useful discussions.

\bibliographystyle{apsrev4-2}
\bibliography{refs}

\end{document}